\documentclass[a4paper,twocolumn,superscriptaddress,showkeys]{revtex4-2}

\pdfoutput=1
\usepackage[utf8]{inputenc}
\usepackage{graphicx,bm,bbm}
\usepackage{amssymb,graphicx,epstopdf}
\usepackage{slashed,subfigure}
\usepackage{caption}
\usepackage{xcolor}
\usepackage{amsmath,mathtools}
\usepackage{epstopdf,dcolumn}
\usepackage{soul}
\usepackage{mathtools}
\allowdisplaybreaks
\usepackage[normalem]{ulem}
\usepackage{cancel}
\usepackage{xcolor}
\usepackage[colorlinks=true,linktocpage=true,linkcolor=blue,citecolor=blue]{hyperref}
\usepackage{xcolor,epsfig,amsmath,amssymb,subfigure,placeins}
\usepackage{graphicx,amsmath,bbm,bm,array}

\newcommand{\be}{\begin{equation}}
	\newcommand{\ee}{\end{equation}}
\newcommand{\ba}{\begin{eqnarray}}
	\newcommand{\ea}{\end{eqnarray}}
\newcommand{\nn}{\nonumber\\}
\newcommand{\ts}{\hspace{.3mm}}

\begin{document}
	\title{Heavy quarkonia in QGP medium in an arbitrary magnetic field}	
	\author{Jobin Sebastian}
	\email{jobin.sebastian@niser.ac.in}
	\affiliation{School of Physical Sciences, National Institute of Science Education and Research,
		An OCC of Homi Bhabha National Institute, Jatni-752050, India}
	
	\author{Lata Thakur}
	\email{thakurphyom@gmail.com}
	\affiliation{Asia Pacific Center for Theoretical Physics,
		Pohang, Gyeongbuk 37673, Republic of Korea}
	\affiliation{Department of Physics and Institute of Physics and Applied Physics, Yonsei University, Seoul 03722, Korea}
	
	\author{Hiranmaya Mishra}
	\email{hiranmaya@niser.ac.in}
	\affiliation{School of Physical Sciences, National Institute of Science Education and Research,
		An OCC of Homi Bhabha National Institute, Jatni-752050, India}
	
	\author{Najmul Haque}
	\email{nhaque@niser.ac.in}
	\affiliation{School of Physical Sciences, National Institute of Science Education and Research,
		An OCC of Homi Bhabha National Institute, Jatni-752050, India}

	\begin{abstract}
		We compute the heavy quarkonium complex potential in a magnetic field of arbitrary strength generated in the relativistic heavy-ion collision. First, the one-loop gluon polarization tensor is obtained in the presence of an external, constant, and homogeneous magnetic field using the Schwinger proper time formalism in Euclidean space. The gluon propagator is computed from the gluon polarization tensor, and it is used to calculate the dielectric permittivity in the presence of the magnetic field. The modified dielectric permittivity is then used to compute the heavy quarkonium complex potential. We find that the heavy quarkonium complex potential is anisotropic in nature, which depends on the angle between the quark-antiquark ($Q\bar{Q}$) dipole axis and the direction of the magnetic field. We discuss the effect of the magnetic field strength and the angular orientation of the dipole on the heavy quarkonium potential. We discuss how the magnetic field influences the thermal widths of quarkonium states. Further, we also discuss the limitation of the strong-field approximation as done in literature in the light of heavy-ion observables, as the effect of the magnetic field is very nominal to the quarkonium potential. 
	\end{abstract}

	\maketitle

	\section{Introduction}
	The heavy quarkonia, which are the bound states of heavy quark and its antiquark  ($c\bar{c},b\bar{b}$) are one of the first proposed signals to probe the deconfining properties of the strongly interacting matter known as quark-gluon plasma (QGP) produced in the heavy ion collision. Matsui and Satz~\cite{Matsui:1986dk} proposed several decades ago that heavy quarkonium production would be suppressed due to the shrinking of the Debye sphere for color interactions in the QGP medium. The presence of various nonequilibrium effects requires the modification of phenomenological models, which can be used to study the properties of the QGP medium. 
	The presence of the nonequilibrium effects such as  momentum space anisotropy  due to viscous effects~\cite{Thakur:2012eb,Thakur:2013nia,Jamal:2018mog,Agotiya:2016bqr,Du:2016wdx, Thakur:2020ifi, Thakur:2021vbo,Islam:2022qmj,Dong:2022mbo,Kurian:2020orp,Singh:2023smw}, %bulk viscousity ~\cite{}, 
	moving medium~\cite{Thakur:2016cki, Escobedo:2011ie,Escobedo:2013tca,Sebastian:2022sga,Chakraborty:2012dt,Patra:2015qoa}, and magnetic field~\cite{ALICE:2015mzu,Marasinghe:2011bt,Alford:2013jva,Machado:2013rta,Cho:2014exa,Guo:2015nsa,Yoshida:2016xgm,Sadofyev:2015hxa,Bonati:2017uvz,Bonati:2015dka, Rougemont:2014efa,Dudal:2014jfa,Karmakar:2018aig,Bandyopadhyay:2017cle,Jamal:2023ncn,Karmakar:2022one,Gowthama:2020ghl} can affect the screening phenomenon, which results in in-medium modification of quarkonium properties.  
	
	In the past decade, the properties of strongly interacting matter have attracted much interest in the presence of magnetic field backgrounds. The noncentral heavy ion collision experiments at RHIC and LHC can produce a very strong magnetic field normal to the reaction plane~\cite{Kharzeev:2007jp,Tuchin:2013ie,Skokov:2009qp,Deng:2012pc,Bzdak:2011yy}, which has motivated several interesting phenomenological studies. These studies led to various novel phenomena such as magnetic catalysis~\cite{Shovkovy:2012zn,Bruckmann:2013oba,Chatterjee:2011ry}, chiral magnetic effect~\cite{Bali:2013esa,Voronyuk:2011jd,Fukushima:2008xe,Alver:2010gr,Kurian:2018qwb,Ghosh:2021knc}, splitting of open charm directed flow~\cite{STAR:2019clv,Das:2016cwd,ALICE:2019sgg,Gowthama:2020ghl}, and modification in properties of heavy quarkonia and dynamics~\cite{Singh:2017nfa,Ghosh:2022sxi,Hasan:2017fmf,Hasan:2020iwa,Zhang:2023ked,Huang:2022fgq,Zhao:2020jqu,Mishra:2020kts,Chen:2021nxs,Iwasaki:2021nrz,Fukushima:2015wck,Kurian:2020kct,Kurian:2019nna,Nilima:2022tmz,Debnath:2023dhs}.
	The potential models have been quite successful in describing the quarkonium properties both in a vacuum as well as in medium~\cite{Lucha:1991vn,Brambilla:2004jw,Karsch:1987pv,Srivastava:2018vxp}. The quarkonium states  are well described by the Cornell potential, which contains both the perturbative Coulombic 
	and nonperurbative confining terms~\cite{Eichten:1974af, Eichten:1979ms}. The emergence of an imaginary part of the potential in the presence of the medium~\cite{Laine:2006ns,Laine:2007qy,Burnier:2007qm,Beraudo:2007ky,Brambilla:2008cx, Brambilla:2011sg, Brambilla:2013dpa} has  instigated the study of heavy quarkonium complex potential~\cite{Dumitru:2009fy,Singh:2017nfa,Margotta:2011ta,Thakur:2013nia,Rothkopf:2011db,Burnier:2012az,Burnier:2015nsa,Rothkopf:2019ipj}. The real part of the quarkonium potential in the magnetic field background has been studied by lattice QCD  in the vacuum and at finite temperature~\cite{Bonati:2018uwh,Bonati:2016kxj}. To our knowledge, no lattice  QCD study has been conducted on the imaginary part of the heavy quarkonium potential. 
	
	In the present work, we aim to study the heavy quarkonium complex potential in the presence of a general magnetic field without any restriction on its strength. The heavy quarkonium potential has been studied previously in the presence of weak and strong magnetic fields based on the assumption that the potential exhibits isotropic behavior~\cite{Singh:2017nfa,Hasan:2017fmf,Hasan:2020iwa}. Recently,  the effect of a general magnetic field on the modification of the imaginary part of the potential has been computed in Ref.~\cite{Ghosh:2022sxi} by considering all Landau level summations and the general structure of the gluon propagator in the magnetic field background. In Ref.~\cite{Ghosh:2022sxi}, the authors have shown that the imaginary part of the potential exhibits anisotropic behavior, but they have not discussed the real part. In this work, we compute both the real and the imaginary parts of the complex heavy quarkonium potential in a constant magnetic field of arbitrary strength. It would be essential to study the effect of the magnetic field on the heavy quarkonium complex potential by assuming the fact that the magnetic field generated in the heavy ion collisions may not be weak or strong compared to the temperature. Here, we employ the Schwinger proper time formalism to study the effect of an external constant magnetic field of arbitrary strength on both the real as well as imaginary parts of the potential. We also discuss the effect of an arbitrary magnetic field on the thermal widths of heavy quarkonium states.
	
	In the current computation, we obtain the in-medium heavy quarkonium complex potential by modifying the Cornell potential with the dielectric permittivity which encodes the effects of the magnetized thermal medium~\cite{Thakur:2012eb,Thakur:2013nia, Agotiya:2008ie}.  The dielectric permittivity is computed using the in-medium gluon propagators. The gluon propagator is obtained from the one-loop polarization tensor in Euclidean space in an external constant and homogeneous magnetic field. The computation is done using the  Schwinger proper time formalism by considering the full interaction between the quark and external field~\cite{Alexandre:2000jc}. Using the magnetic field-modified dielectric permittivity, we obtain the heavy quarkonium complex potential in an arbitrary magnetic field. We find that the potential obtained exhibits anisotropic behavior because of the quark loop contribution to the gluon self-energy. We demonstrate the effects of a general magnetic field on the complex heavy quarkonium potential and decay width obtained using the imaginary part of the $ Q\bar{Q} $ potential and check the validity of strong field approximation in the potential model of quarkonia.

	This paper is structured as follows. In Sec.~\ref{sec:SelfE}, we revisit the derivation of the gluon polarization tensor in the presence of an arbitrary magnetic field. We obtain the gluon propagator in the static limit and compute the dielectric permittivity. In Sec.~\ref{sec:HQpot}, we compute the heavy quarkonium complex potential and discuss the effect of the magnetic field on it. In Sec.~\ref{sec:Dwidth}, we estimate the quarkonium state's thermal width and discuss how the magnetic field affects them. In Sec.~\ref{approximation}, the strong field approximation of polarization tensor is computed, and the corresponding effect on the heavy quarkonium potential is studied and compared with an arbitrary magnetic field scenario. We summarize our results in Sec.~\ref{summary}.
	%%%%%%%%%%%%%%%%%%%%%%%%%%%%%%%%%%%%%%%%

	%%%%%%%%%%%%%%%%%%%%%%%%%%%%%%%%%%%%%%%%
	\section{Dielectric permittivity}
	\label{sec:SelfE}
	In this section, we derive the dielectric permittivity in the presence of an arbitrary magnetic field, which we use later to compute the in-medium heavy quarkonium potential. First, we calculate the gluon self-energies and propagators in an arbitrary magnetic field.
	\subsection{Gluon self-energy in an arbitrary magnetic field}
	In the following,  we review the computation of the longitudinal component of gluon self-energy in  the one-loop order as followed in Ref.~\cite{Alexandre:2000jc} and obtain the longitudinal component of the gluon self-energy and propagator in the static limit.
	%Debye screening mass and gluon propagator in the static limit. 
	Consider a  charged particle of charge $ q_f $ and mass $ m $  moving in an external, constant, and homogeneous magnetic field, which is directed along the z direction ($ {\bf B}=B\hat{z} $).  Here we choose the symmetric gauge; therefore, we have
	%\begin{eqnarray}
	%	{A_0}^{ext}(x)=0, 	{A_1}^{ext}(x)=-\frac{B}{2}y, \nonumber\\
	%		{A_2}^{ext}(x)=+\frac{B}{2}x, 	{A_3}^{ext}(x)=0, 
	%\end{eqnarray}
	\begin{eqnarray}
		{A_0}(x)=0, \,\,\,\,\,	{A_1}(x)=-\frac{B}{2}y, \nonumber\\
		{A_2}(x)=\frac{B}{2}x, \,\,\,\,\,	{A_3}(x)=0.
		\label{symgauge}
	\end{eqnarray}
	In coordinate space, the fermion propagator, as introduced by Schwinger, is given by~\cite{Schwinger:1951nm}
	\begin{equation}
		S(x,x')=e^{i e x^\mu A_\mu (x')}\widetilde{S}(x-x'),
	\end{equation}
	where the phase factor $e^{i e x^\mu A_\mu (x')}  $ does not contribute to the gluon self-energy with this particular choice of the gauge~(\ref{symgauge}). The Fourier transform  $\widetilde{S}(k)$ of the translational invariant part of fermion propagator $\widetilde{S}(x-x')$ in the proper time formalism is 
	\begin{eqnarray}
		\widetilde{S}(k)&=&\int_{0}^{\infty}ds\, e^{is[k_0^2-k_3^2-k^2_\perp\tan(|q_fB|s)/|q_fB|s-m^2]}\nonumber\\
		&\times&\{(k_0\gamma_0-k_3\gamma_3+m)[1+\gamma_1\gamma_2 \tan(|q_fB|s)]\nonumber\\
		&-&k_\perp \gamma_\perp[1+\tan^2(|q_fB|s)]\},
	\end{eqnarray}
	where $ k_\perp=(k_1,k_2) $ is the transverse momentum.  For the finite temperature case, we note that the bosonic Matsubara modes $ \omega_n=2 n \pi T $ and fermionic ones $ \hat \omega_l=(2 l+1) \pi T$, respectively. The fermion propagator, in Euclidean space $ (k_0=i\hat \omega_l) $ along with $ s\rightarrow -is $ as followed in Ref.~\cite{Alexandre:2000jc}, becomes
	\begin{eqnarray}
		\widetilde{S}_l(\bf{k})&=&-i\int_{0}^{\infty}ds\, e^{-s[\hat \omega_l^2+k_3^2+k^2_\perp\tanh(|q_fB|s)/|q_fB|s+m^2]}\nonumber\\
		&\times&\{(-\hat \omega_l\gamma_4-k_3\gamma_3+m)[1-i\gamma_1\gamma_2 \tanh(|q_fB|s)]\nonumber\\
		&-&k_\perp \gamma_\perp[1-\tanh^2(|q_fB|s)]\},
		\label{Fermion_prop}
	\end{eqnarray}
	where $ {\bf{k}} =(k_\perp,k_3)$ and $ \gamma_{\mu} $ $(\mu=1,2,3,4)$ are the 
	Euclidean gamma matrices, which satisfy the anticommutation relation $ \{\gamma_\mu,\gamma_\nu\}=-2\delta_{\mu\nu} $.
	
	Based on the fermion propagator~(\ref{Fermion_prop}), one can obtain the quark-loop contribution to the one-loop gluon self-energy in the presence of a magnetic field as
	\begin{eqnarray}
		\Pi_n^{\mu \nu}({\bf{p}}, B)&=&-g^2T\sum_f\int \frac{d^3{\bf k}}{(2\pi)^3}\sum_{l=- \infty}^{\infty} {\rm tr}\{\gamma^\mu \widetilde{S}_{l}({\bf k})\nonumber\\
		&\times& \gamma^{\nu}\widetilde{S}_{l-n}({\bf{k-p}})\}+Q^{\mu\nu}( p),
	\end{eqnarray}
	where, with $ p=|{\bf p|} $, $ Q^{\mu\nu}(p) $ is the ``contact" term, which cancels the ultraviolet divergences and is independent of both the temperature and magnetic field. Here, we are interested in the longitudinal component of the gluon polarization tensor, which we will use further to compute the dielectric permittivity and hence the in-medium heavy quarkonium potential. 
	The longitudinal component of the quark contribution to the one-loop gluon self-energy is obtained after integration over $ {\bf k} $~\cite{Alexandre:2000jc}.
	%%%%%%%%%%%%%
	\begin{widetext}
		\begin{eqnarray}
			\Pi^{44}_{n,q} (\omega_n, {\bf{p}},  B)&=&-\sum_{f}\frac{g^2 T}{8\pi^{3/2}}q_fB\int\limits_{0}^{\infty}du\  u^{1/2}\int\limits_{-1}^{1}dv\sum_{l=- \infty}^{\infty}\exp \left[\frac{p_\perp^2}{q_fB} \frac{\cosh(  q_fBuv)-\cosh (q_fBu)}{ 2 \sinh q_fBu} \right.\nn
			&&\left.- u\left\{m^2+W_l^2
			+\frac{1}{4}(1-v^2)(\omega_n^2+p_3^2)\right\}\right]\left[\frac{p_\perp^2}{2}\frac{ \cosh (q_fBuv)-v \coth ( q_fBu)\, \sinh( q_fBuv)}{\sinh( q_fBu) }\right. \nn
			&&\left.\hspace{1cm}-\frac{1 }{u} \coth (q_fBu) \Big(1-2\,u\,W_l^2
			+\frac{1}{2}u\,v\,\omega_n W_l-u(1-v^2)p_3^2\Big)\right]
			+Q^{44}(p),
		\end{eqnarray}	
	\end{widetext}
	where $ W_l= \hat \omega_l-[(1-v)/2]\omega_n$. The contact term $ Q^{44}(p) $ is independent of temperature and magnetic field and hence can be obtained as both $ T $ and $ B $ approach zero. \\

	After using the Poisson resummation, one can isolate the temperature-independent and temperature-dependent parts from the longitudinal polarisation tensor ~\cite{Alexandre:2000jc}. As we aim to study the effect of the magnetic field on the heavy quark-antiquark potential in the medium, we only discuss the temperature-dependent part of the gluon self-energy. The temperature-dependent part of the longitudinal polarisation tensor in the limit of massless quarks becomes
	\begin{widetext}
		\begin{eqnarray}
			\Pi^{44}_{n,q} (\omega_n,{\bf p},  B) &=& -\sum_{f}\frac{ g^2}{(4 \pi )^2} q_fB\int_{0}^{\infty}du \int_{-1}^{1}dv\sum_{l\geq 1}(-1)^l\exp \Big[-\frac{p_\perp^2}{q_fB} \frac{\cosh  (q_fBu)-\cosh( q_fBuv)}{ 2 \sinh (q_fB\, u)}-\frac{1}{4} u \left(1-v^2\right)\nonumber\\ 
			&&\times \left(p_3^2+\omega_n^2\right)\Big] 
			e^{ -\frac{l^2}{4 T^2 u}} \left[\cos (\pi  l n (1-v)) \left\{p_\perp^2  \frac{\cosh( q_fBuv )-v \coth(  q_fBu) \, \sinh (q_fBuv)}{\sinh  (q_fBu) } \right. \right.\nonumber\\
			&&+\text{p}_3^2 \left(1-v^2\right)\coth( q_fBu)\bigg\} 
			-\frac{\coth (q_fBu) }{u} \left\{\frac{l^2 }{T^2 u}\cos \pi  l n (1-v)-2 \pi \, l\, n\, v \sin \pi  l n (1-v)\right\} \bigg].
			\label{PiLq}
		\end{eqnarray}
		After evaluating the sum over $ l $, the above Eq.~(\ref{PiLq}) in the static limit becomes %($  \omega_{n}\rightarrow 0, n\rightarrow 0$)  as
		%\begin{widetext}
		\begin{eqnarray}
			\Pi_q^{44} (0, {\bf p},B) &=&\Pi_q^{L}({\bf p},B),\nonumber\\
			&=& \sum_{f}\frac{ g^2}{32\pi^2} q_fB\int_{0}^{\infty}du \int_{-1}^{1}dv\exp \left[-\frac{1}{4}p_3^2 u(1-v^2)-\frac{ p_{\perp}^{2}}{q_fB}\frac{\cosh( q_fBu) -\cosh( q_fBuv)}{ 2\sinh(q_fBu)}\right]\nonumber\\
			&\times& \bigg[ 4\coth q_fBu\ \frac{\partial}{\partial u}\vartheta _4 (0,e^{-\frac{1}{4 T^{2}u}}) %e^{-\frac{1}{4 T^{2}u}}\frac{1}{2 T^{2}u^2}
			 + \left(1-\vartheta _4\left(0,e^{-\frac{1}{4 T^2 u}}\right)\right)
		\nonumber\\
			&&\hspace{1cm} \times	\left\{p_{\perp}^{2} \frac{\cosh (q_f Buv)}{\sinh(q_fBu)}
		+
			\coth (q_fBu)\left(p_3^2-p_3^2v^{2}-p_{\perp}^{2}v~\frac{\sinh (q_fBuv )}{\sinh (q_fBu)}\right) \right\}\bigg],
			\label{PiLqw0}
		\end{eqnarray}
		where $\vartheta _4$  is the Jacobi Theta function and obtained as
		\be
		\sum_{l=1}^\infty (-1)^l e^{-a l^2} =\frac{ 1}{2}\left[ \vartheta_4(0,e^{-a})-1\right].
		\ee
		In spherical polar coordinates, Eq.~\eqref{PiLqw0} becomes
		\begin{eqnarray}
			\Pi_q^{L}({\bf p},B) &=&\sum_{f}\frac{ g^2q_f B}{32\pi^2} \int_{0}^{\infty}du \int_{-1}^{1}dv\exp \left[-\frac{1 }{4}p^2\cos^2\theta\, u(1-v^2)-\frac{ p^{2}\sin^2\theta}{2\, q_fB \sinh(q_fBu)}
			(\cosh (q_fBu)-\cosh (q_fBu v))\right] \nn
			&\times&\left[ 4\coth (q_fBu)\ \frac{\partial}{\partial u}
			\vartheta _4 (0,e^{-\frac{1}{4 T^{2}u}})
		%	e^{-\frac{1}{4 T^{2}u}}\frac{1}{2 T^{2}u^2} 
			+ \left(1-\vartheta _4(0,e^{-\frac{1}{4 T^2 u}})\right)
			\bigg\{p^{2}\sin^2\theta \cosh (q_fBuv){\rm csch}(q_fBu)
			\right.\nonumber\\
			&&\left.+
			\coth (q_fBu)\left(p^2 \cos^2 \theta(1-v^2)-v\,p^{2}\sin^2\theta~\frac{\sinh( q_fB uv)}{\sinh(q_fBu)} \right)\bigg\} \right].
			\label{PiLqwsph}
		\end{eqnarray}
	\end{widetext}
	%%%%%%%
	Here the coupling constant $ g $ depends upon magnetic field, i.e., $g^2(\Lambda^2,B)
	%[\equiv g^2(\Lambda^2,B)]
	=4 \pi \alpha_s(\Lambda^2,B)$,
	where $ \alpha_s$ is  the one-loop running coupling constant in the magnetic field background as followed in~\cite{Ayala:2018wux,Bandyopadhyay:2017cle}
	\begin{equation}
		\alpha_{s}(\Lambda^{2},|eB|)=\frac{\alpha_{s}(\Lambda^{2})}{1+b_{1}\alpha_{s}(\Lambda^{2})\ln\left(\frac{\Lambda^{2}}{\Lambda^{2}+|eB|}\right)},
	\end{equation}
	and the one-loop strong coupling in the absence of any magnetic field is
	\begin{equation}
		\alpha_{s}(\Lambda^{2})=\frac{1}{b_{1}\ln\left(\frac{\Lambda^{2}}{\Lambda_{\overline{\rm MS}}^{2}}\right)},
		\label{alpha}
	\end{equation}
	where $ b_{1}=\frac{(11N_{c}-2 N_{f})}{12 \pi} $ and $\Lambda_{\overline{\rm MS}}=176 $ MeV for $ N_f=3 $. Here we  take $ \Lambda$ for quarks as $\Lambda_q=2\pi\sqrt{T^2+\mu^2/\pi^2} $  and for gluons as $ \Lambda_g=2\pi T $. We take the zero chemical potential $ (\mu) $ here. %in the present computation.\\
	The quark loop contribution to the  gluon self-energy, $ \Pi_q^{L} ( {\bf p}) $ for  $ B=0 $ case can be written as
	\begin{eqnarray}
		\Pi_q^{L} ( {\bf p})&=&-\frac{3\, g^2}{2\, \pi^2}\int_{0}^{\infty} \frac{k\, dk}{e^{k/T}+1}\nn
		&\times&\left[2+\frac{(p^2-4k^2)}{2 \,k\, p} {\rm \log}\left(\frac{p-2\,k}{p+2\,k}\right)\right].
	\end{eqnarray}	
	The magnetic field dependence only comes through the quark loop contribution to the gluon self-energy, as gluons do not interact with the magnetic field. Therefore, the gluon contribution to the self-energy remains the same as without the magnetic field.
	%%%%
	\begin{equation}
		\Pi_g^{L}(\omega, {\bf p})%&=&\Pi_g^{L}(\omega, {\bf p})\nonumber\\
		=m_{Dg}^2\bigg[1-\frac{\omega}{2 p}\ln\bigg(\frac{\omega+p}{\omega-p}\bigg)+i\pi\frac{\omega}{2 p}\Theta(p^2-\omega^2)\bigg],
		\label{Pig}
	\end{equation}
	where  $m_{Dg}^2=\frac{g'^2 T^2 N_c}{3}$ with 
	$ g'^2=4\pi \alpha_s(\Lambda^2) $ with $\alpha_s(\Lambda^2) $ defined in Eq.~(\ref{alpha}). 
	\\
	The  above equation (\ref{Pig}) can be rewritten in terms of real and imaginary parts as
	\begin{eqnarray}
		\Re\Pi_g^{L}(\omega, {\bf p})&=&m_{Dg}^2\bigg[1-\frac{\omega}{2 p}\ln\bigg(\frac{\omega+p}{\omega-p}\bigg)\bigg],\nonumber\\
		\Im\Pi_g^{L}(\omega, {\bf p})
		&=&m_{Dg}^2\frac{\pi\omega}{2 p}\Theta(p^2-\omega^2).
	\end{eqnarray}
	The total longitudinal component of gluon self-energy is the sum of the gluon and quark contribution 
	\begin{equation}
		\Pi^{L} (\omega_n,{\bf p},B)  = \Pi_g^{L} ( \omega_n, {\bf p}) +	\Pi_q^{L} ( \omega_n,{\bf p},B) ,
	\end{equation}	
	which can be written in terms of real and imaginary parts. We compute the gluon self-energy's real and imaginary parts in the static limit ($ \omega \rightarrow 0 $). The real part of self-energy reads
	%The  gluon self-energy can be written in terms of the real and imaginary parts; the real part of self-energy reads
	\begin{equation}
		\Re\Pi^{L} (\omega,{\bf p},B)  =	\Re\Pi_g^{L} (\omega, {\bf p}) +	\Re\Pi_q^{L} ( \omega,{\bf p},B),
	\end{equation}	
	and the imaginary part of the self-energy $\Im \Pi^{L} $  reads
	\begin{eqnarray}
		&&\Im\Pi^{L} (\omega,{\bf p},B) 
		=\Im\Pi_g^{L} ( \omega, {\bf p})	+	\Im\Pi_q^{L} ( \omega,{\bf p},B).
	\end{eqnarray}
	%%%%%%%
	The imaginary contribution from the quark loop can be obtained by using the identity
	\begin{eqnarray}
		&&\hspace{-0.5cm}\Im \Pi^L_{n,q}(\omega_n,{\bf p})\nn
		&=&\frac{1}{2i}\lim\limits_{\varepsilon\rightarrow0}\bigg[\Pi^L_q(\omega_n+i\varepsilon,{\bf p})-\Pi^L_q(\omega_n-i\varepsilon,{\bf p})\bigg].
		\label{img}
	\end{eqnarray}
	Further, we compute both the real and imaginary part of the longitudinal component of the gluon propagator using the gluon self-energy.
The spectral function approach, as defined in Ref.~\cite{Weldon:1990iw}, is used to obtain the imaginary part of the gluon propagator as
	\begin{equation}
		\Im{D^{L}}(\omega,{\bf p})=-\pi(1+e^{-\beta \omega})\rm A^{L},
		\label{prop}
	\end{equation}
	where $\rm A^{L}$ is defined as
	\begin{equation}
		\rm A^{L}(\omega,{\bf p})=\frac{1}{\pi}\frac{e^{\beta \omega}}{e^{\beta \omega}-1}\rho^L(\omega,{\bf p}).
	\end{equation}	
	The spectral function  $\rho^L$  can be expressed in the Breit-Wigner form  as
	\ba
	&&\rho^L(\omega_n,{\bf p},B) \nn
	&=&\frac{\Im{\Pi^L}(\omega_n,{\bf p},B)}{( p^2-\Re{\Pi^L}(\omega_n,{\bf p},B))^2+{\Im \Pi^L(\omega_n,{\bf p},B)}^2},
	\label{spctrl}~~~~~
	\ea
	where $p=|{\bm p}|$. After substituting Eq.~(\ref{spctrl}) in Eq.~(\ref{prop}), we obtain the longitudinal component of the gluon propagator, $ D^{L} $ 
	in terms of real and imaginary parts. In the static  $ (\omega\rightarrow 0) $ and massless light quark limit, we obtain 
	\begin{equation}
		D^{L}({\bf p},B)=\frac{-1}{p^2+	\Pi^{L} ({\bf p},B) }+\frac{i\pi T\, \Pi^{L}({\bf p},B) }{p(p^2+	{\Pi^{L} ({\bf p},B) })^2}.
	\end{equation}
	Using the gluon propagator, we obtain the dielectric permittivity as~\cite{Thakur:2012eb, Thakur:2013nia, Singh:2017nfa}
	\begin{equation}
		\epsilon^{-1}({\bf p},B)=\frac{p^2}{p^2+\Pi^{L}}-i\pi T\frac{p\,    \Pi^{L}}{(p^2+\Pi^{L})^2},
		\label{eps}
	\end{equation}
	where $ \Pi^{L}\equiv \Pi^{L} ({\bf p},B)$.\\
	%%%%%%%%%%%%%%Figure1%%%%
	
	We use the dielectric permittivity expression~\eqref{eps} to compute the in-medium heavy quarkonium complex potential in an arbitrary magnetic field.

\section{In-medium heavy quarkonium  potential }%in the presence of magnetic field}
\label{sec:HQpot}
In this section, we obtain the in-medium heavy quarkonium potential by using the modified dielectric permittivity computed in the previous section. We obtain the in-medium heavy quarkonium potential %in the coordinate space ($ V({\bm r})$) 
by correcting the Cornell potential in Fourier space with the dielectric permittivity, which encodes the effects of the magnetized thermal medium~\cite{Agotiya:2008ie,  Thakur:2013nia, Thakur:2016cki} 
\ba
V({\bm r, T, B})=\int \frac{d^3{\bm p}}{(2\pi)^{3/2}} ~(e^{i{\bm {p\cdot r}}}-1)~\frac{ V_{\text{Cornell}}(p)}{\epsilon(  p,B)},\label{vmod}
\ea
where $ V_{\rm Cornell}(p) $ is the Fourier transform of the Cornell potential $ V_{\rm Cornell}(r)=-\alpha/r +\sigma \,r $, which is given by
% Here the Cornell potential 
% defined as
% \begin{equation}
	% 	V_{\rm Cornell}(r)=-\frac{\alpha}{r}+\sigma \,r,
	% \end{equation}
%and is given by
\begin{equation} 
	V_{\rm Cornell}(p)=\sqrt{\frac{2}{\pi}}\frac{\alpha}{p^2} -2\sqrt{\frac{2}{\pi}}\frac{\sigma}{p^4},
	\label{Vp}
\end{equation}
where $ \alpha$ and $ \sigma $ are the strong coupling constant and the string tension, respectively.
%
%%%%%%%%%%figure%%%%%%%%%
%%%%%%%%%%%Figurerealpartpot%%%%%%%%%%%%%
\begin{figure*}[tbh]
	\centering
	\includegraphics[width=8cm]{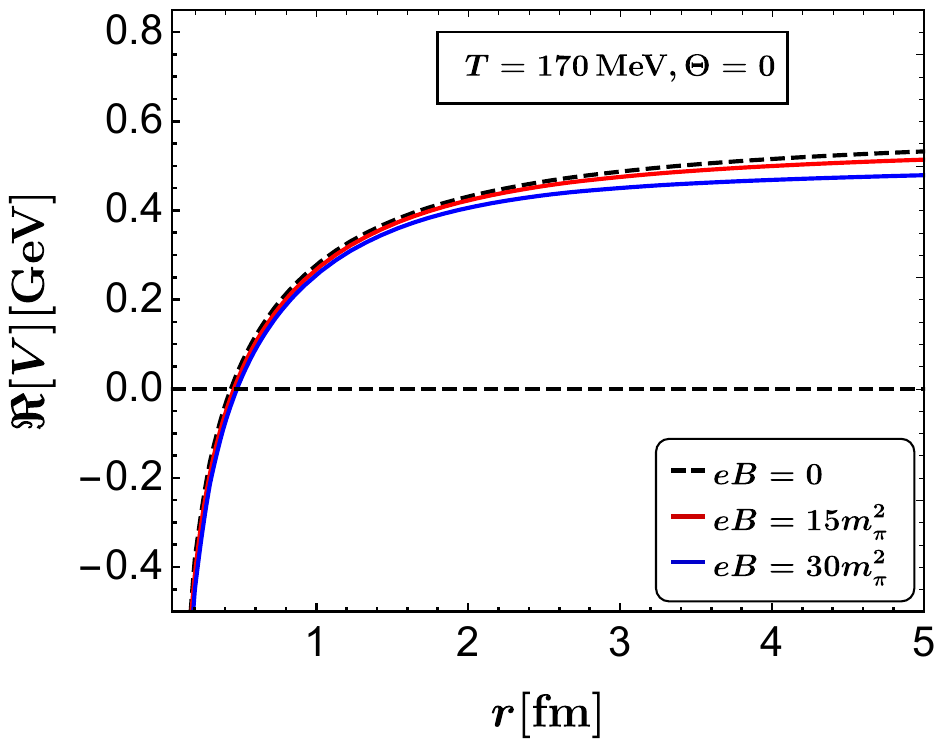}
	\hspace{10mm}
	\includegraphics[width=8cm]{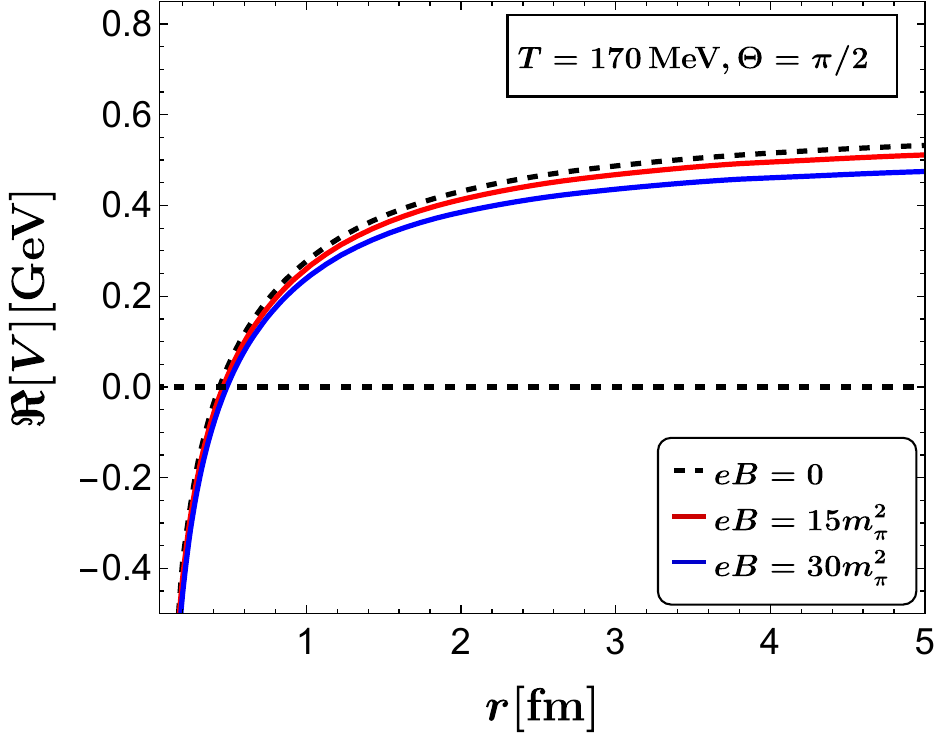}%[scale=0.36]{imr2.pdf}
	\caption{The real part of the potential is plotted as a function of quark-antiquark separation $r$ for $\Theta=0$ (left) and $\Theta=\pi/2$ (right)
	at $ T=170 $~MeV. } %The left panel shows the variation of the real part of the potential. The right panel shows the variation of the imaginary part of the potential.}
	\label{repot}
\end{figure*}
%
%%%%%%%%figure%%%%%%%%%%%%%%
\begin{figure*}[tbh]
	\centering
	\includegraphics[width=8cm]{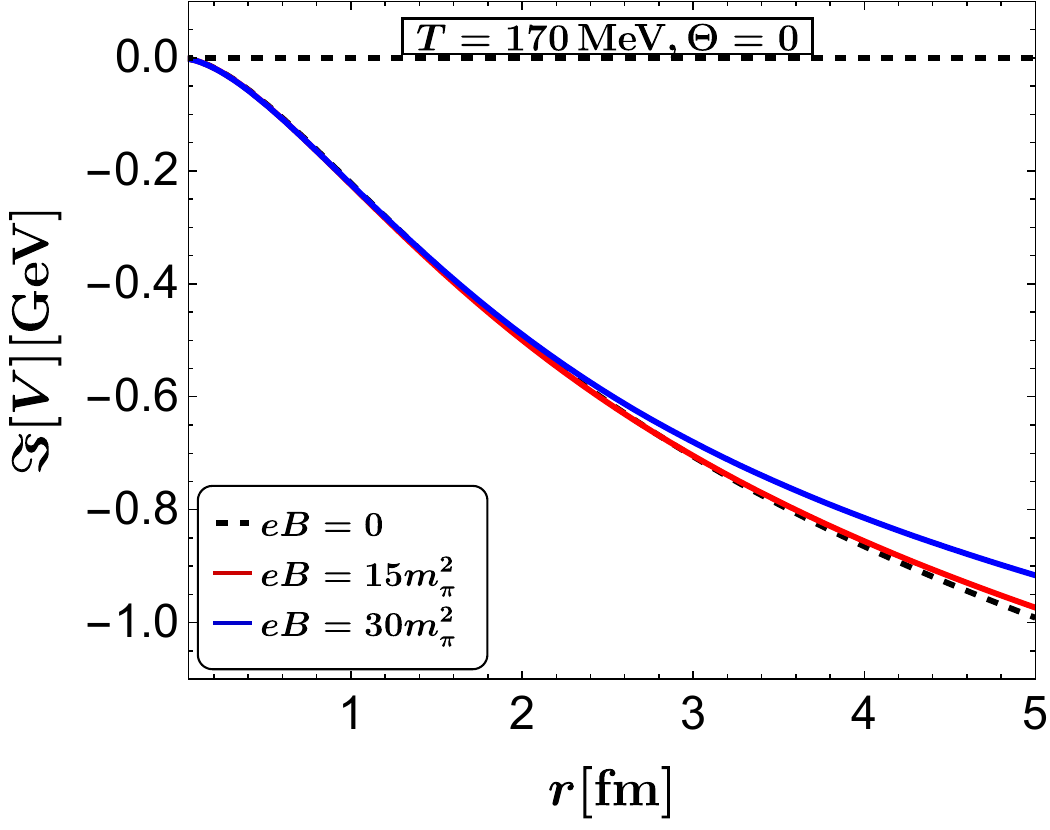}
	\hspace{10mm}
	\includegraphics[width=8cm]{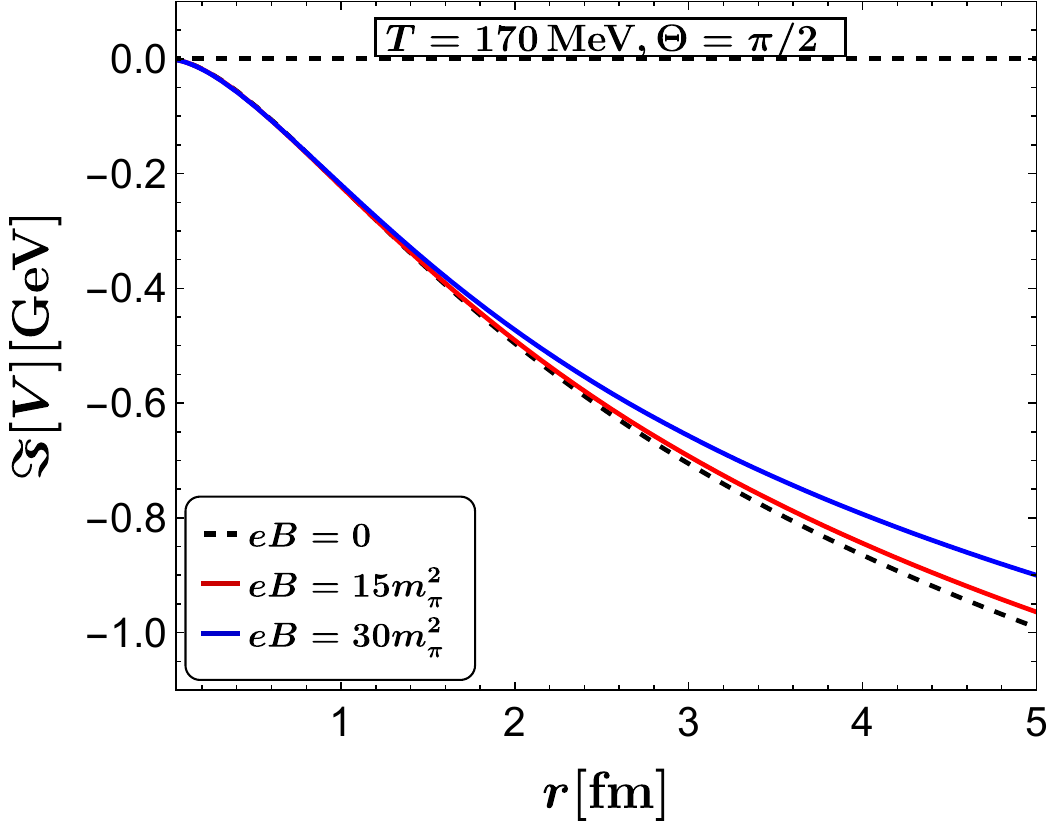}%[scale=0.36]{imr2.pdf}
	\caption{The imaginary part of the potential is plotted as a function of quark-antiquark separation $r$ for $\Theta=0$ (left) and $\Theta=\pi/2$ (right) 
		at $ T=170 $~MeV.} % The left panel shows the variation of the real part of the potential. The right panel shows the variation of the imaginary part of the potential.}
	\label{Impot}
\end{figure*}
%
%%%%%%%figure%%%%%%%%%%%%%%
\begin{figure*}[tbh]
	\centering
	\includegraphics[width=8cm]{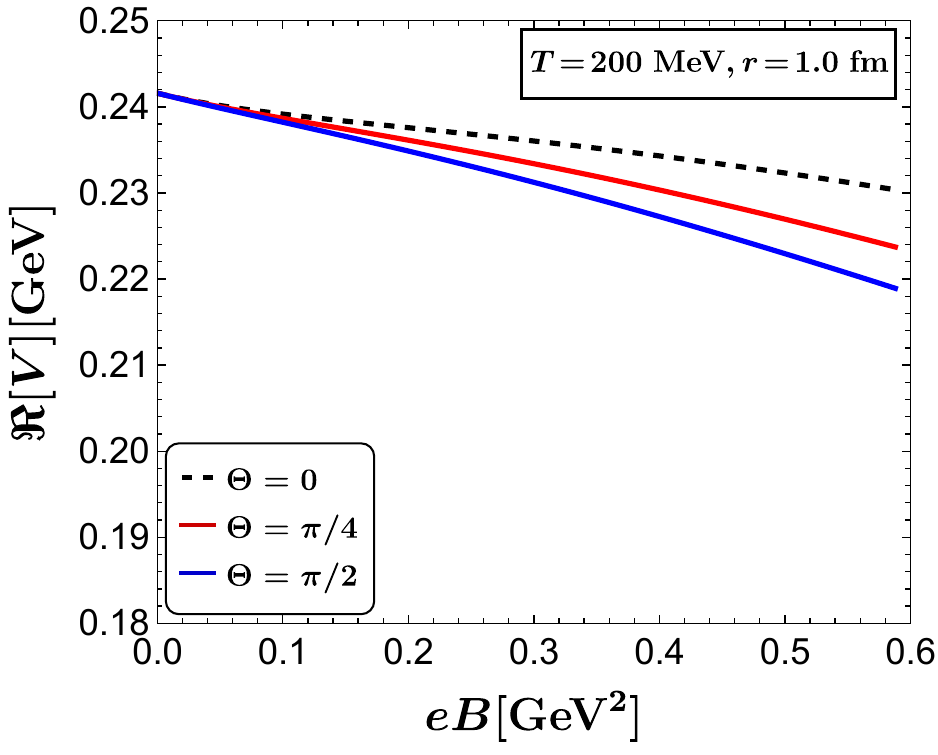}
	\hspace{10mm}
	\includegraphics[width=8.5cm]{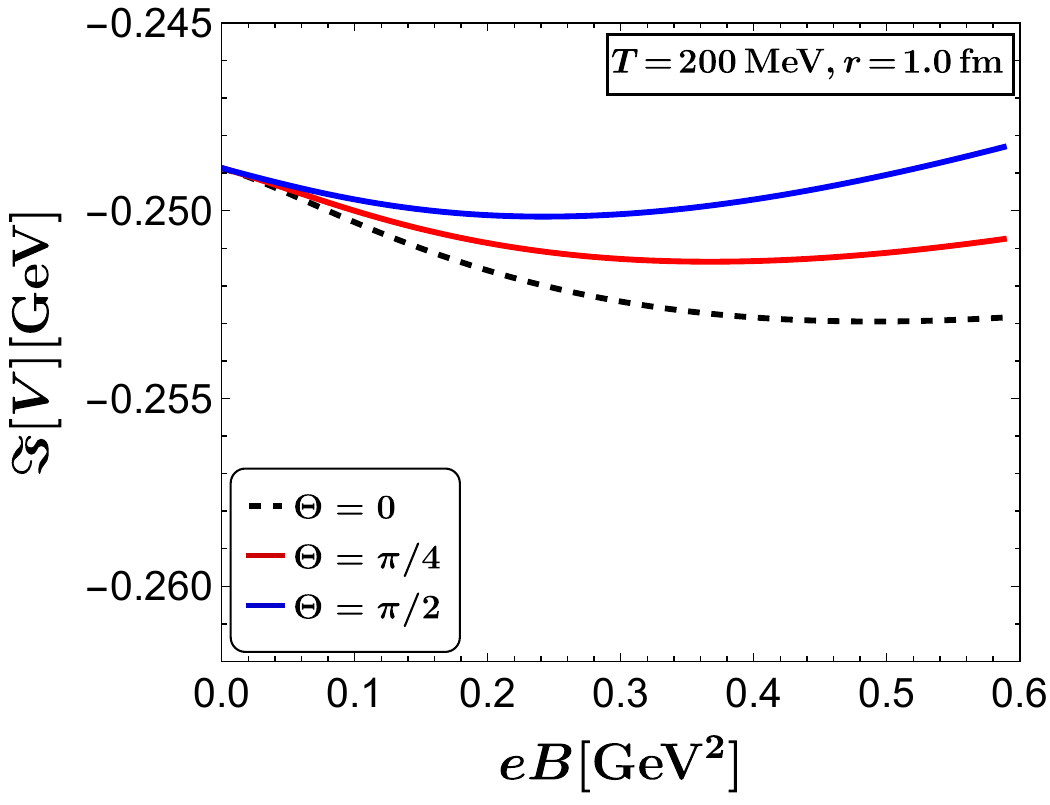}%[scale=0.36]{imr2.pdf}
	\caption{The complex potential is plotted as a function of the strength of the magnetic field for different values of $\Theta$ when $ r=1  $ fm and $ T=200 $~MeV. The left panel shows the variation of the real part of the potential. The right panel shows the variation of the imaginary part of the potential.}
	\label{re_impot}
\end{figure*}
%%%%
Here we take $ \alpha=4/3 \,\alpha_{s} (\Lambda^2, B)$ and $ \sigma=0.18 $ GeV$ ^2 $ and
$\epsilon(p)$ is the dielectric permittivity, which is defined in Eq.~(\ref{eps}).  
After substituting Eqs.~(\ref{eps}) and (\ref{Vp}) in Eq.~(\ref{vmod}), we obtain both the real as well as imaginary part of the potential, 	which contains both the perturbative Coulombic and nonperturbative string terms. The real part of the potential can be written in terms of Coulombic and string terms as
\ba 
\Re V({\bm r,T,B})=\Re V_c({\bm r, T,B})+\Re V_\sigma({\bm r, T, B}),
\ea 
where the Coulombic term is  
\ba
\Re V_c(\bm r, T, B)&=&-\frac{\alpha}{2\pi^2} \int d^3{\bm p} \nn
&\times&\Bigg[\frac{e^{i\boldsymbol{p\cdot r}}}{p^2 +\Pi^{L}}
-\frac{\Pi^{L}}{p^2(p^2 +\Pi^{L})}\Bigg],~~~~~
\ea
and the string term reads %$ \Re V_\sigma({\bm r}) $ is
\ba
\Re V_\sigma(\bm r, T, B) = -\frac{\sigma}{\pi^2} \int\!\!  \frac{d^3{\bm p}}{(2\pi)^3}(e^{i{\bm p}\cdot {\bm r}}-1)\frac{1}{p^2(p^2 +\Pi^{L})}.\hspace{0.65cm}
\ea
Here
%\ba 
${\bm p} \cdot {\bm r}=r p \big[\sin \theta  \sin \Theta  \cos (\phi -\Phi )+\cos \theta \cos \Theta\big]$
%\ea
and the angles $\theta (\Theta)$ and $\phi(\Phi)$ are polar and azimuthal angles in momentum (coordinate) space, respectively. 
After integrating over the azimuthal angle, we obtain
%%%%%
\ba
\Re V({r, T, B, \Theta} )=-\frac{1}{\pi}\int \frac{\sin\theta\, d\theta\, dp}{p^2+\Pi^{L}} \Big[(\alpha \,\Pi^{L}-2\,\sigma)\nonumber\\ 
\hspace{-1.5cm}\times(\alpha\, p^2+2\,\sigma)e^{i p  r  \cos \theta \cos \Theta }J_0(p\ts r\ts \sin\theta  \sin \Theta )\Big],
\label{eq:Repot}
\ea
where $J_0$ is the Bessel's function of the first kind. 
%\vspace{1cm}
%%%%%%%%%%%%%%%%%%%%%%

Similarly, we compute the imaginary part of the quarkonium potential. The imaginary part of the potential is given by 
%%%%%%%%%
\ba
\Im V({\bm r, T, B})= T\int \frac{d^3{\bm p}}{2\pi}\frac{(e^{i\bm{p\cdot r}}-1) \Pi^{L} p}{\left(p^2 + \Pi^{L}\right)^2} \left[\frac{\alpha}{p^2} +\frac{2 \sigma}{p^4}\right].~~~~
\ea
After integrating over the azimuthal angle, we obtain
\ba
\Im V({r, \Theta, T, B})=-T\int\frac{\sin\theta\, d\theta\, dp}{\left(p^2+\Pi^{L}\right)^2}\Pi^{L}\left[
\alpha\ts p +\frac{2\sigma }{p}\right]\nonumber\\
\times\Big\{1-e^{i p r \cos \theta \cos \Theta} J_0( p r \sin \theta \sin \Theta )\Big\}.\,\,\,\,\,\,\,\,
\label{eq:Impot}
\ea
%%%%%%%%%%%%%%%%%%%%%%%%%%%%%%
%%%%%%%%%%%%%%%%%%%%%figure
\begin{figure*}[tbh]
	\centering
	\includegraphics[width=8cm]{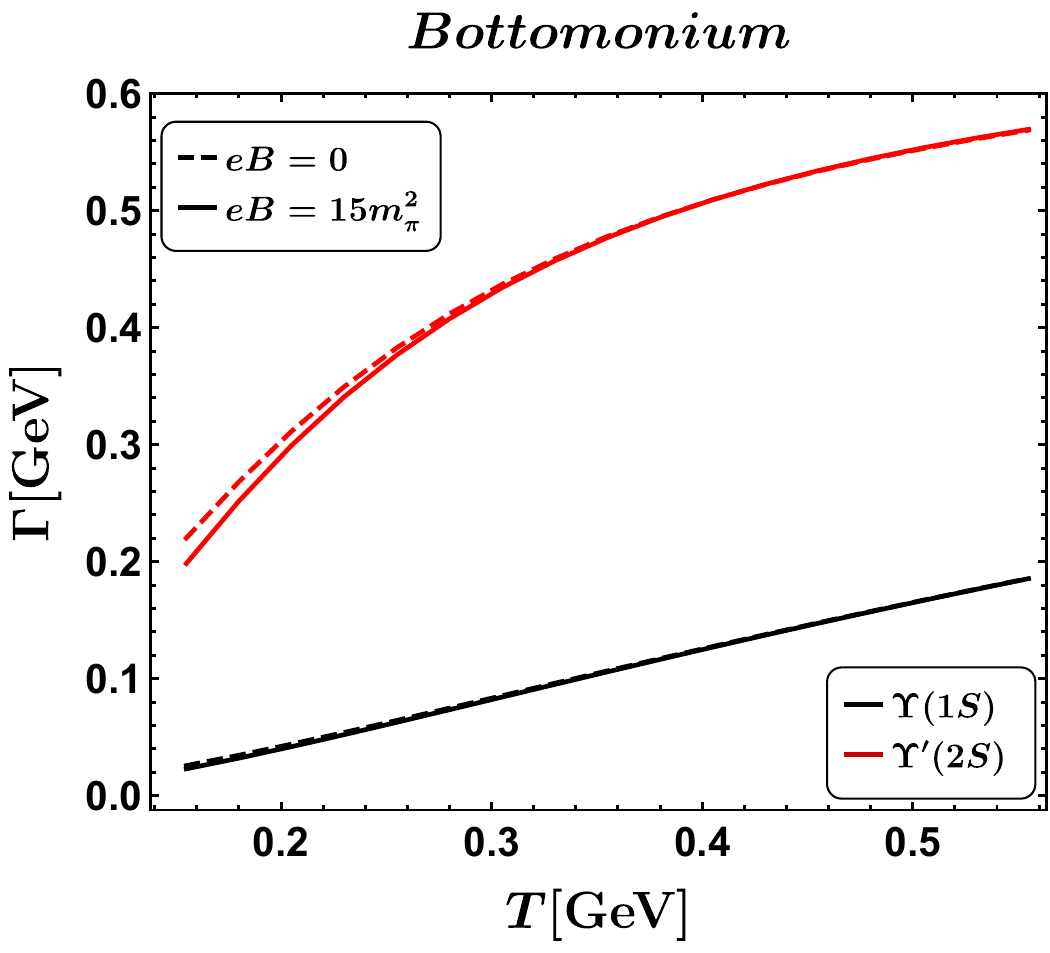}
	\hspace{10mm}
	\includegraphics[width=8cm]{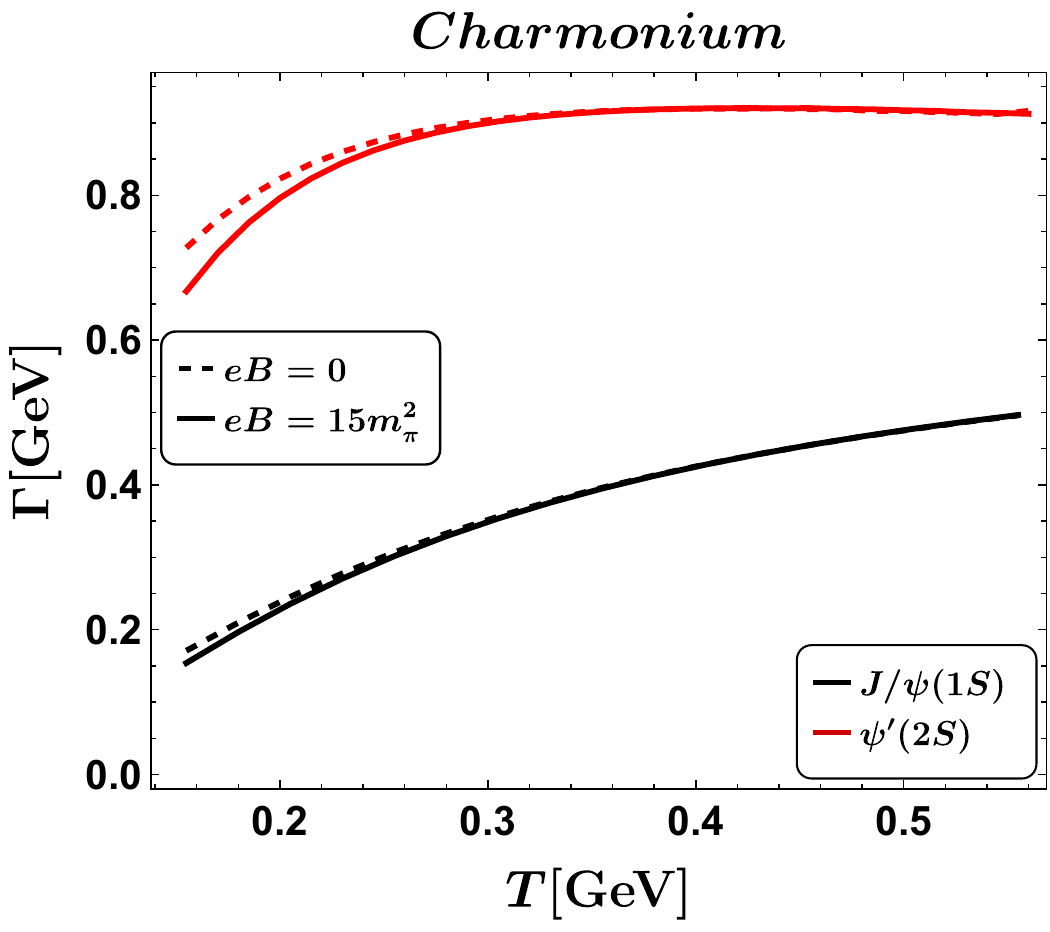}%[scale=0.36]{imr2.pdf}
	\caption{ The thermal widths of bottomonium (left) and charmonium (right) states as a function of temperature at $ B=0$ and $15 m_{\pi}^2$ are shown.}
\label{DW1}
\end{figure*}
We numerically solve the real (\ref{eq:Repot}) and imaginary (\ref{eq:Impot}) parts of the potential. 
In Fig.~\ref{repot}, we plot the real part of the potential as a function of separation distance $ r $ for different strengths of the magnetic field $ eB $. 
The left panel shows  for the $ Q\bar{Q} $ dipole axis alignment  along the direction of the magnetic field ($ \Theta=0 $), whereas the right panel shows its perpendicular alignment with respect to the magnetic field ($\Theta=\pi/2 $).
%
%The left panel shows for $ \Theta=0 $, when
%%for two  different orientation of 
%$ Q\bar{Q} $ is aligned along  the direction of  magnetic field   and the right panel shows  for the perpendicular alignment of $ Q\bar{Q} $ concerning the direction of the magnetic field ($ \Theta=\pi/2 $). 
We find that the real part of the potential becomes flattened with the magnetic field, due to an increase in screening with $ B $. The effect of screening is seen to be slightly higher in the perpendicular case than along the direction of magnetic field. 

In Fig.~\ref{Impot}, we plot the
 imaginary  part of the potential for $ \Theta=0 $ (left) and $ \Theta=\pi/2 $ (right) for the different values of the magnetic field.  The imaginary part of the potential shows different behavior at smaller and larger $ r $; it increases with the magnetic field at smaller $ r $ and decreases in magnitude with the increase in the magnetic field at larger $ r $.  
 The decrease in magnitude with the magnetic field is observed to be higher for $ \Theta=\pi/2 $ compared to $ \Theta=0 $.
 Note that the magnetic field dependence is insignificant for the potential, especially in the range  $ eB=0$ to $eB=15 m_{\pi}^2 $.
 
%In figure~\ref{repot}, we plot the real (left) and imaginary (right) part of the potential for different strengths of the magnetic field $ B $ at $ \Theta=0 $.  We find that the real part of the potential becomes flattened with the magnetic field due to an increase in screening with $ B $.	The imaginary part of the potential shows different behavior at smaller and larger $ r $; it increases with the magnetic field at smaller $ r $ and	decreases in magnitude with the increase in the magnetic field at larger $ r $.  Note that the magnetic field dependence is nonsignificant for the potential, especially in the range  $ B=0$ to $B=15 m_{\pi}^2 $  %and is more suppressed at larger $ r $.
%

Figure~\ref{re_impot} shows the real and imaginary part of the potential as a function of magnetic field for different values of $ \Theta $ at $ r =1 $ fm. We observe that the real part of the potential varies in response to a magnetic field at different rates according to direction. The magnetic field dependence is found to be maximum in $\Theta=\pi/2$ direction and minimum along the direction of the magnetic field, which establishes the anisotropy of the potential in the magnetic field. The magnitude of the imaginary part of the potential initially increases when the magnetic field increases, and diminishes as the magnetic field intensifies. Both components of the potential depend on both the magnitude of the magnetic field and the angle, however, this dependence is minimal.

The imaginary part increases in magnitude with the increase in the magnetic field initially, and the magnitude decreases as the magnetic field increases. Both the parts of the potential depend on the magnetic field and the angle between the dipole axis and the magnetic field, but the dependence is rather weak.

In the next section, we use the imaginary part of the potential to obtain the thermal widths of the quarkonium states.    
%
%%%% section %%%%%%%%%%%%%%%%%%%%%%
%%%%%%%%%%%%%%%%%%%%%%%%%%%%%%%%%%%%
\section{Thermal Width }
\label{sec:Dwidth}
%%%
\begin{figure*}[tbh]
\centering
\includegraphics[width=8.2cm]{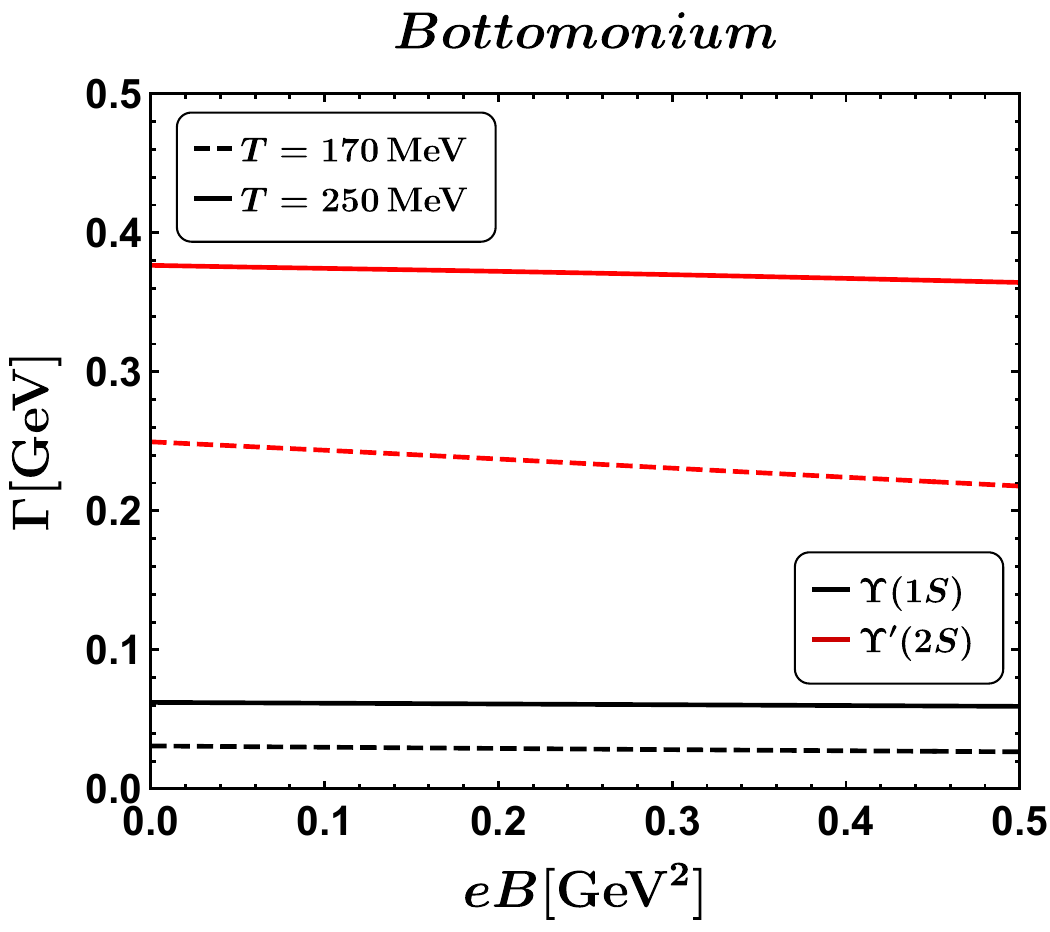}
\hspace{10mm}
\includegraphics[width=8cm]{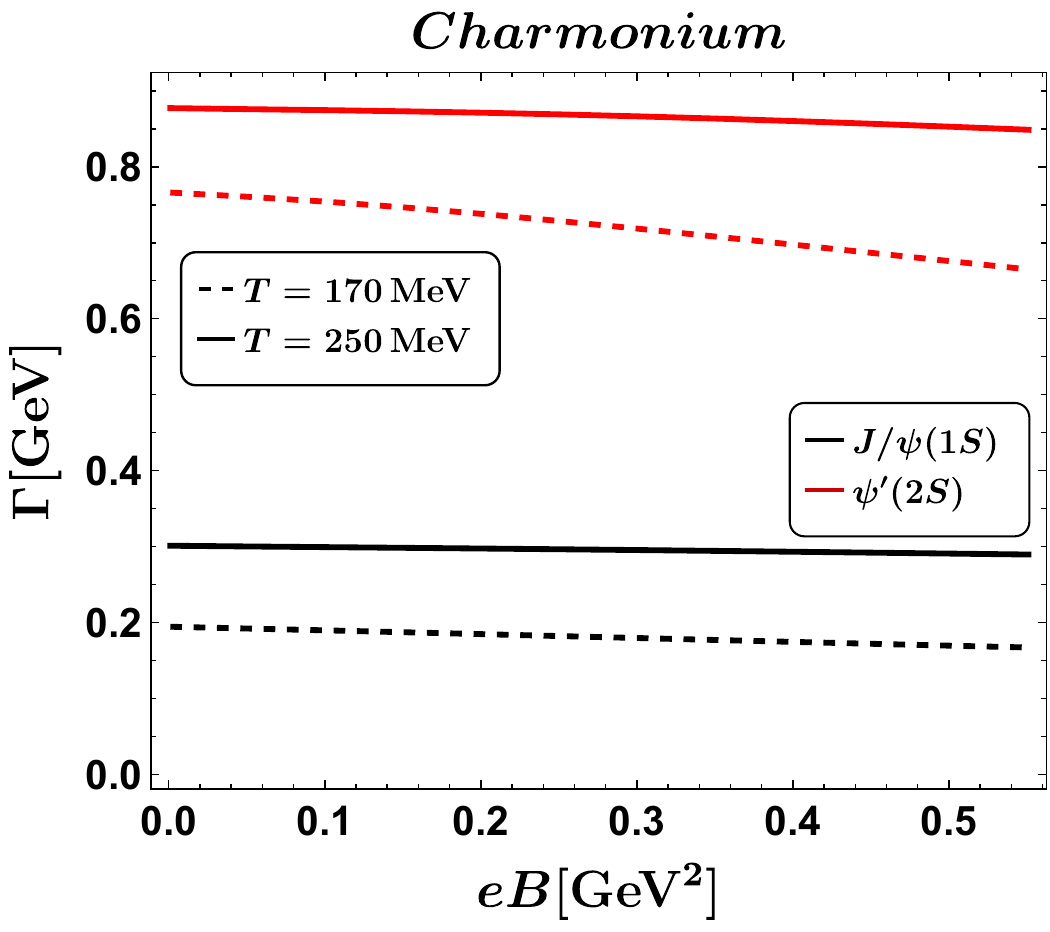}%[scale=0.36]{imr2.pdf}
\caption{Thermal widths of bottomonium (left) and charmonium (right) states as a function of magnetic field for $ T=170$~MeV  and $250$~MeV. }%Imaginary part of the potential as a function of separation distance $ r $ for $ B=0, 15 m_{\pi}^2, 30 m_{\pi}^2 $ at $ T=170 $~MeV with $ \Theta=0 $ (Left) and $ \Theta=\pi/2 $ (Right)}
\label{DW2}
\end{figure*}
%%%%%%%%%%%%%%%%%%%%%
In this section, we compute the thermal widths of the quarkonium states. The imaginary part of the potential estimates the thermal width, $\Gamma_{Q\bar{Q}}$, when treated as a perturbation of the vacuum potential. 	We compute the  decay width of the quarkonium states as~\cite{Thakur:2013nia, Dumitru:2009fy}	
%%%%%%%
\begin{equation}
\Gamma_{Q\bar{Q}}(T, B) = - \langle \psi(r) | \Im \, V_{Q\bar{Q}}(r,T, B, \Theta) | \psi (r)\rangle,
\label{eq:thermalwidth}
\end{equation}
where $ \psi (r)$ is the unperturbed Coulombic wave function. As the leading contribution to the imaginary potential for a deeply heavy-quark bound state is Coulombic, which justifies the use of Coulomb wave functions to calculate the thermal width. The wave function for the ground and excited states is given by
\begin{eqnarray}
\psi_{1s}({r}) &= &\frac{1}{(\pi a_0^{3})^{1/2}} e^{-r/a_0}, \nonumber\\
\psi_{2s}({r})& =& \frac{1}{4(2\pi a_0^{3})^{1/2}}  \left(2-\frac{r}{a_{0}}\right)e^{-r/2 a_0},
\label{eq:wavefunction}
\end{eqnarray}
\\
where $ a_{0} =2/C_{F}m_{Q}\alpha_{s}$ 	is the Bohr radius of the $ Q\bar{Q} $ system and $m_Q$ is the heavy quark mass.  After substituting Eqs.~(\ref{eq:wavefunction}) and (\ref{eq:Impot}) in Eq.~(\ref{eq:thermalwidth}), we obtain the thermal width of the quarkonium states for the ground state as
\begin{equation}
\label{eq:thermalwidth1S}
\Gamma_{1s}(T, B) %&=&-\frac{1}{\pi a_0^3} \int d^3{\bf r} \,  e^{-2r/a_0}\, 
%\Im \,V_{Q\bar{Q}}(r,T) \nonumber\\
=-\frac{1}{\pi a_0^3} \int d^3{\bf r} \,  e^{-2r/a_0}
\Im V\left(r,T, B,\Theta\right),\, 
%	\Gamma_{1s} &=& \Gamma_{\alpha(1s)} + \Gamma_{\sigma(1s)}   
\end{equation}

\begin{widetext}
\begin{eqnarray}
\Gamma_{1s}(T, B)&\!\!=\!\!&\frac{2 T}{a_0^3}\int dr \, d\Theta \, r^2 \sin\Theta e^{-2r/a_0} \int \frac{\sin\theta\, d\theta\, dp}{\left(p^2+\Pi^{L}\right)^2}
\Pi^{L}\left[
\alpha\ts p+\frac{2 \sigma }{ p}\right],\hspace{.6cm}
\label{eq:dw_1s}
\end{eqnarray}
and for the first excited states ($ 2S $) as
\begin{eqnarray}
\Gamma_{2s}(T, B)=\frac{T}{16 a_0^3}\int dr \, d\Theta \, r^2 \sin\Theta \left(2-\frac{r}{a_{0}}\right)^2 e^{-r/a_0} \int \frac{\sin\theta\, d\theta\, dp}{\left(p^2+\Pi^{L}\right)^2}
\Pi^{L}\left[
\alpha\ts p+\frac{2 \sigma }{ p}\right].\hspace{.6cm}
\label{eq:dw_2s}
\end{eqnarray}
\end{widetext}
We numerically compute the thermal widths of the ground~(\ref{eq:dw_1s}) and excited~(\ref{eq:dw_2s}) states of the bottomonium and charmonium states.  In Fig.~\ref{DW1}, we present plots of the thermal widths of the ground and first excited states of bottomonium (on the left) and charmonium (on the right) as a function of temperature for both $eB=0$ and $eB=15 m_\pi^2$.

It is observed that the thermal width increases with an increase in temperature, as anticipated. The magnetic field effect is more on the first excited state than the ground state for both the bottomonium and charmonium states. At larger $ r $, the effect of the magnetic field on the imaginary part of the potential is more due to the larger size of the excited states. Hence the excited states are more sensitive to the magnetic field than the ground state. Figure~\ref{DW2} shows the variation of thermal width with the magnetic field at different temperatures for bottomonium (left) and charmonium (right) states. We find that the thermal widths are more sensitive to the magnetic field at lower temperatures than the higher temperatures. The magnetic field effects decrease with the increase in heavy quark mass and decrease in the size of bound states. It can be concluded from the figures that the magnetic field has only a negligible effect on the thermal width compared to the temperature.

%%%%%%%%%%%%%%%%%%%%%%	

%%%%%%%%%%%%%%%%%%%%%%%%%%%%%%%%%%%%%%%%%%%%%%%%%%%%%%%%%%%%%%%%%%%%%%%%%%%%%%%%%%
\section{Strong  field approximation}
\label{approximation}
%%%%%%%%%%%%%%%%%%%%%%%%%%%%%%%%%%%%%%%%%%%%%%%%%%%%%%%%%%%%%%%%%%%
\begin{figure*}[tbh]
\centering
\includegraphics[width=8cm]{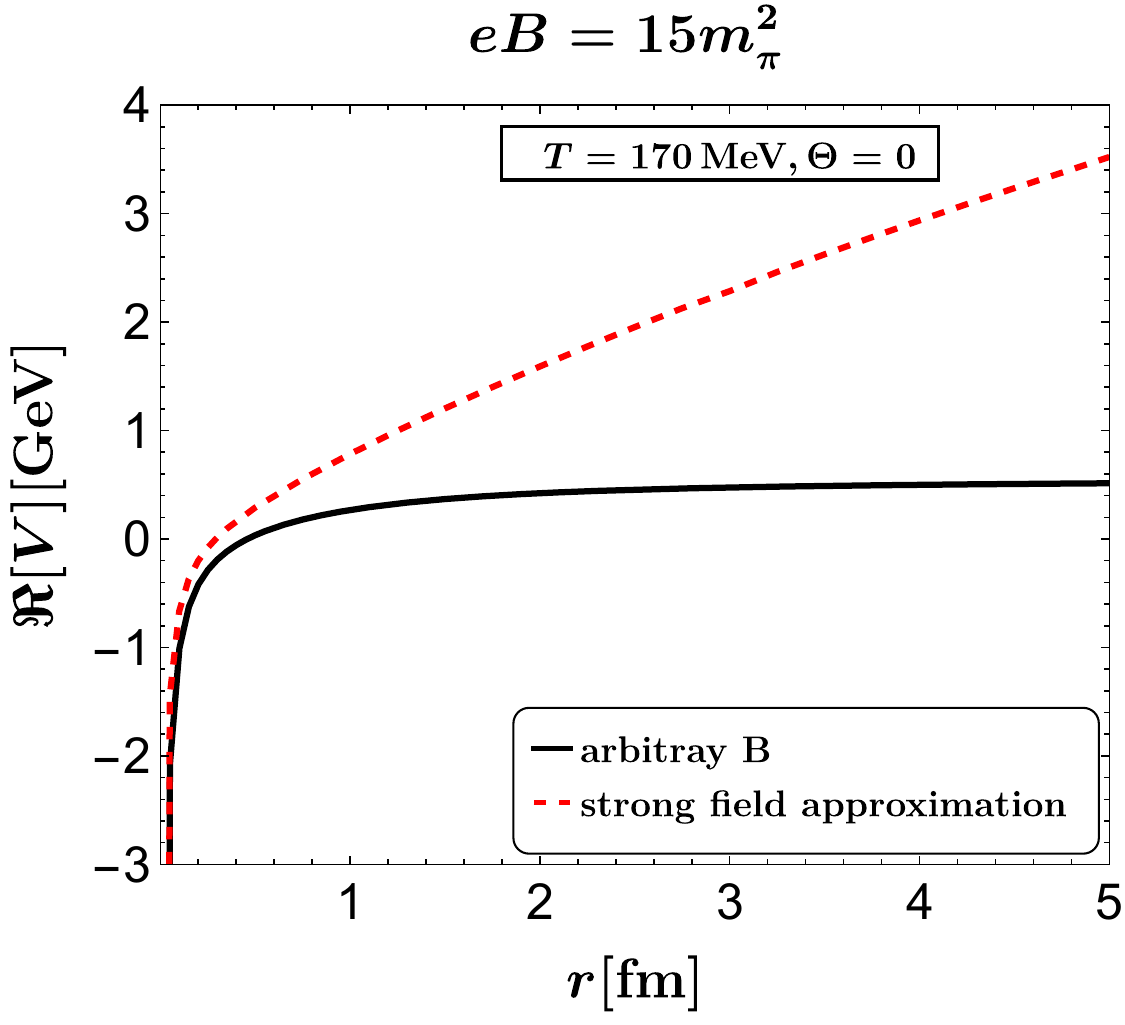}
\hspace{10mm}
\includegraphics[width=8cm]{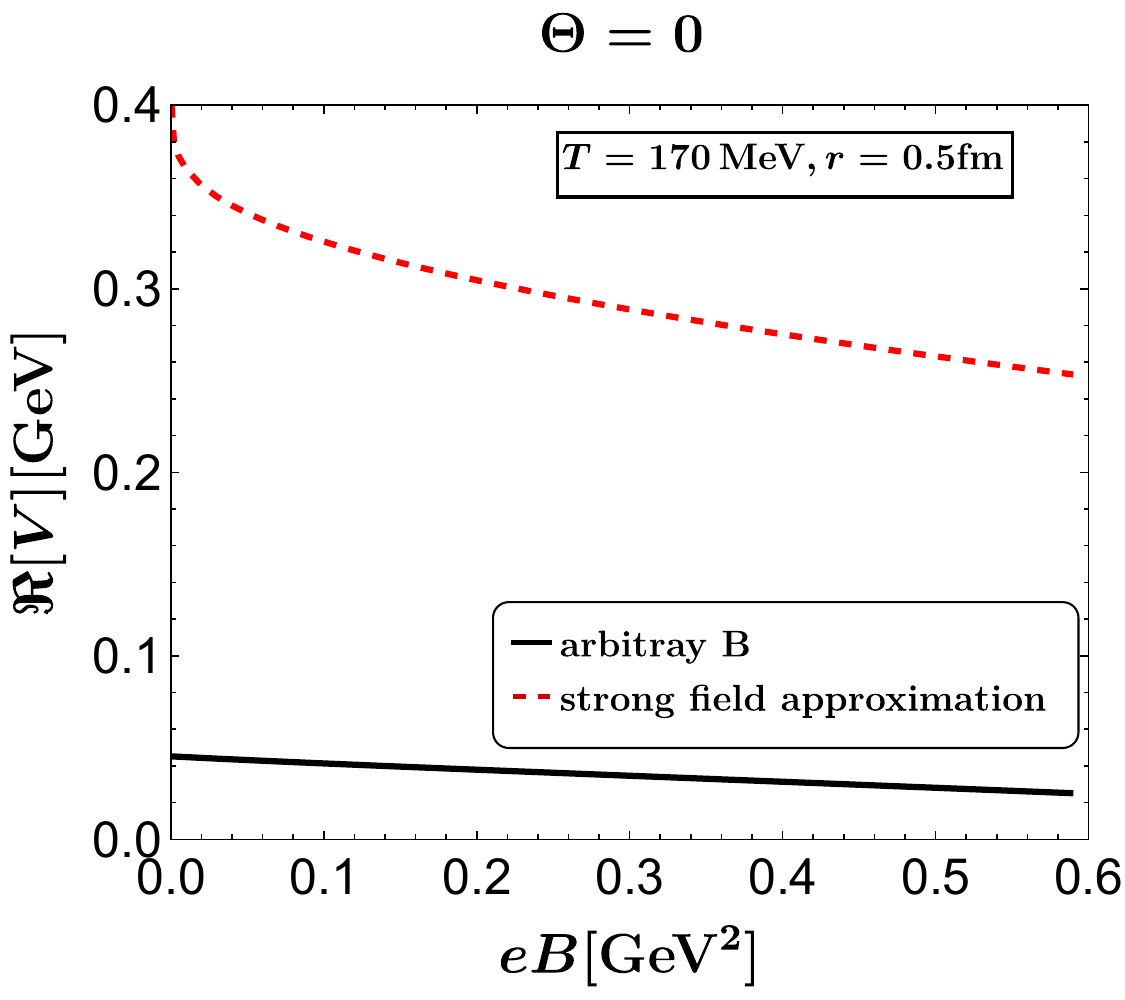}%[scale=0.36]{imr2.pdf}
\caption{The real part of the potential as a function of $r$ is shown on the left side for an arbitrary (black solid) and strong magnetic field approximation (red dashed) at $T = 170$~MeV and $\Theta = 0$. The right side illustrates the same potential as a function of $eB$ at these conditions.
}
\label{Fig:ReVcompare}
\end{figure*}
%%%%%%%%%%%%%%%%%%%%%%%%%%%%%%%%%%%%%%%%%%%%%%%%%%%%%%%%%%%%%%%%%%%
In this section, we compute the longitudinal component of the gluon self-energy	in the strong magnetic field approximation ($  T \ll \sqrt{|eB|}$). In the strong magnetic field limit  ($ |eB|\rightarrow\infty $), the fermion propagator [Eq.~(\ref{Fermion_prop})] for massless case becomes
\begin{eqnarray}
\widetilde{S}_l(\bf{k})&=&-i\int_{0}^{\infty}ds\, e^{-s[\hat \omega_l^2+k_3^2+k^2_\perp/|q_fB|s]}\nonumber\\
&\times&(-\hat \omega_l\gamma_4-k_3\gamma_3)[1-i\gamma_1\gamma_2], \nonumber\\
&=&ie^{-k^2_\perp/|q_f B|}\frac{\hat \omega_l\gamma_4+k_3\gamma_3}{\hat \omega_l^2+k_3^2}[1-i\gamma_1\gamma_2],
\end{eqnarray}
which is similar to the expression for the fermion propagator computed for the lowest Landau level approximation in Refs.~\cite{Gusynin:1995nb,Karmakar:2018aig}.
%%%%%
In $|eB|\rightarrow\infty $ limit, the temperature dependent part of the longitudinal component of  the gluon self-energy~(\ref{PiLq}) reduces to the dominant term as
\begin{eqnarray}
\Pi_{q}^{L}({\omega_n,\bf p},B)&=& -\sum_{f} \frac{ g^2}{(4\pi)^2} \frac{(q_fB)^2}{T^2}e^{-p_{\perp}^{2}/2q_fB}\int_{0}^{\infty}\frac{du}{u^2} %\frac{2 
\nonumber\\
&	\times& \int_{-1}^{1}dv \,e^{-c u }\sum_{l\geq 1}(-1)^{(l+1)}l^2 e^{-d/u}\nonumber\\
&\times&	\cos \pi  l n (1-v),
\label{PiL_Binf}
\end{eqnarray}	
where $ c= (1-v^2)(p_3^2+\omega_n^2)/(4\,q_fB)$ and $ d=l^2q_fB/4T^2 $. The integration over $ u $ can be done analytically using the relation
\begin{equation}
\int_{0}^{\infty}\frac{du}{u^2}e^{-cu-d/u}=2\sqrt{\frac{c}{d}}K_{1}(2\sqrt{c\,d}).
\end{equation}
Here $ K_n (z)$ represents the modified Bessel function of the second kind. Therefore, the longitudinal component of the gluon self-energy (\ref{PiL_Binf}) for the strong magnetic field approximation in the static limit ($ \omega_n\rightarrow 0 $) becomes %equation~
%(\ref{PiL_Binf}) becomes
\begin{eqnarray}
\Pi_{q}^{L}({\bf p},B)&=& -\sum_{f}\frac{ g^2 q_fB}{8\pi^2T} e^{-p_{\perp}^{2}/2 q_fB} \,%\frac{2 
	\sum_{l\geq 1}(-1)^{(l+1)} \nonumber\\
	&&\hspace{-1cm}\times  \int\limits_{-1}^{1}dv\,l\ p_3\sqrt{1-v^2}K_{1}\left(\frac{l\, p_3}{2T}\sqrt{1-v^2}\right). %\cos \pi  l n (1-v). \nonumber\\
	\label{strogB}
\end{eqnarray}
%%%%%%%%
Further, we compute the Debye screening mass from Eq.~(\ref{PiLq}) in the limit $ p\rightarrow 0 $,  for the regime where $  T \ll \sqrt{eB }$  
\begin{eqnarray}
	m_{D}^2(T, B)&=&-\lim_{{\bf p}\to0}\Pi^{L}_{q} ({\bf p},  B), \nonumber\\
	&=& \sum_{f}\frac{ g^2q_fB}{8 \pi ^2T^2} \int_{0}^{\infty}\frac{du}{u^2}\coth q_fB\,u  \sum_{l\geq 1}(-1)^{l+1}l^2\nonumber\\
	&&\vspace{0.5cm}\ \times\ e^{-l^2/4uT^2},\nonumber\\
	&=&\sum_{f}\frac{g^2q_fB}{4\pi^2}.
\end{eqnarray}
%%%%%%%
In the absence of a magnetic field, the Debye screening mass becomes
\begin{eqnarray}
	m_D^2(T)&=& \frac{1}{3}C_Ag^2T^2+
	\sum_f\frac{ g^2}{8\pi^2T^2} \int_{0}^{\infty}\frac{du}{u^3} \nonumber\\ &&\times \sum_{l\geq 1}(-1)^{l+1}l^2 e^{-l^2/4uT^2},\nonumber\\
	&=&\frac{1}{3}C_Ag^2 T^2 +\frac{1}{6}N_fg^2T^2.
\end{eqnarray}
%{\color{red}12 should be 6}.\\
%%%%%%%%%%%%%%%%%%%%
The longitudinal component of gluon self-energy in strong field approximation obtained in Eq.~(\ref{strogB}) can be substituted in Eq.~(\ref{vmod}) to study the behavior of quarkonium potential in the strong field approximation. In Fig.~\ref{Fig:ReVcompare}, we show the effect of arbitrary magnetic fields and the strong field approximation on the real part of the potential. We find that the real part of the potential is more suppressed for the case of an arbitrary magnetic field as compared to strong field approximation due to larger screening in an arbitrary $ B $. We can see that the potential with approximation differs in large values from the exact potential for any realistic magnetic field magnitude, and the difference gradually reduces as the magnetic field increases. Hence we can say that the strong field approximation is invalid, and one should consider the general case while studying the properties of quarkonium states.

\section{Summary }
\label{summary}
In the present work, we have evaluated the influence of a magnetic field on the heavy quarkonium complex potential. We initially computed the dielectric permittivity from the static limit of the gluon propagator. 
%%%%%%%
This propagator was derived from the one-loop gluon self-energy in the presence of an external magnetic field, which was evaluated using Schwinger's proper time formalism in Euclidean space. The effect of the magnetic field enters through the quark-loop contribution to the gluon self-energy and coupling constant. Then, we computed the in-medium heavy quarkonium complex potential using the modified dielectric permittivity. Results showed that this potential is anisotropic and varies with magnetic field strength and angle $\Theta$ between quark-antiquark axis and direction of magnetic field.

For very high magnetic field strengths, the real part of the potential gets flattened due to an increase in screening with $eB$. On the other hand, the imaginary part of the quarkonium potential experiences a rise in magnitude at short distances, followed by a decrease at long distances. Furthermore, the inclusion of a magnetic field introduces an angular dependence into the potential. Finally, we observed that the overall effect of the magnetic field on the complex potential is rather small for realistic strengths of magnetic field generated in heavy ion collisions.

We computed the decay widths of the ground and first excited states of bottomonium ($ \Upsilon,  \Upsilon^{\prime}$) and charmonium ($J/\psi, \psi^{\prime}$) using the imaginary part of the potential. We found that the excited states $ (\Upsilon^{\prime}, \psi^{\prime} )$ are more sensitive to the magnetic field than the ground states $ (\Upsilon, J/\psi) $. The effect of magnetic fields decreases with increasing heavy quark mass and decreasing size, making  the charmonium states more sensitive to magnetic field strength than the bottomonium states. For the decay widths, as the temperature increases, the sensitivity to magnetic fields decreases, eventually disappearing at high temperatures.

We have further compared our results with the strong-field approximated potential. We found that such an approximated potential does not even come close to the potential without such an approximation for any realistic magnetic field value generated in heavy ion collisions. 
%have any significant correction due to any realistic magnetic field generated in heavy ion collision.
%even come close to any realistic magnetic field value generated in the heavy ion collisions. 
The approximation makes the screening much weaker as compared to an estimation for the arbitrary magnetic field. The strong magnetic field approximated screening however gradually increases as the magnetic field increases. The present investigation invalidates %As done in previous works, we invalidate 
the strong magnetic field approximation usually adopted in literature for the heavy quarkonium complex potential. For the realistic strengths of magnetic fields, one needs to take the effects of a  general magnetic field as has been attempted here. Moreover, it may also be noted that, the weak-field expansion introduces new divergences in the gluon propagators, and one needs a way to regulate it. Hence, it is essential to study the effect of arbitrary magnetic fields on the heavy quarkonium complex potential and the properties of quarkonium states. 

In the future, we would like to extend our computation for the moving medium in the presence of an arbitrary magnetic field.

\section{Acknowledgments}
J.~S.~ would like to thank R.~Ghosh for the useful discussions.
L.~T.~ is supported by the Korean Ministry of Education, Science and Technology, Gyeongsangbuk-do and Pohang City at the APCTP
and  National Research Foundation (NRF) funded by the Ministry of Science of Korea (Grant No. 2021R1F1A1061387). N.H. is supported in part by the SERB-MATRICS under Grant No. MTR/2021/000939.

%\providecommand{\href}[2]{#2}\begingroup\raggedright
%\endgroup

\end{document}